# Electrostatically actuated silicon-based nanomechanical switch at room temperature

Diego N. Guerra, Matthias Imboden and Pritiraj Mohanty [a)]

*Department of Physics, Boston University, 590 Commonwealth, Boston, Massachusetts 02215*

**We demonstrate a silicon-based high frequency nanomechanical device capable of switching controllably between two states at room temperature. The device uses a nanomechanical resonator with two distinct states in the hysteretic nonlinear regime. In contrast to prior work, we demonstrate room temperature electrostatic actuation and sensing of the switching device with 100% fidelity by phase modulating the drive signal. This phase-modulated device can be used as a low-power high-speed mechanical switch integrated on-chip with silicon circuitry.**

Silicon-based nanomechanical switching devices at megahertz-range frequencies are of fundamental and technical interests. A suspended nanomechanical structure with two distinct excitation states can be used as an archetypal two-state system to study a plethora of fundamental phenomena such as Duffing nonlinearity,[1] stochastic resonance[2] and macroscopic quantum tunneling at low temperatures.[3,4,5] From a technical perspective, there are numerous applications in which micro- and nanoscale mechanical structures with two distinct fixed states (static) or resonant states (dynamic) can be used as microwave switches.[6] Since the two states can result in high or low capacitance relative to a fixed electrode, such a device can be used as a switch with high-frequency short or open configurations. In addition, a more tantalizing possibility is their on-chip integration with complementary metal-oxide-semiconductor (CMOS) circuitry.[7]

Towards this end, there has been a lot of research activity, such as the development of doubly-clamped nanomechanical beam structures consisting of two distinct states that could serve as a memory element.[8] Although this device demonstrated the fundamental aspect of controlled switching between two states in the hysteretic regime, the experimental realization was carried out at sub-Kelvin temperatures using the magnetomotive measurement scheme. To be viable for commercial applications, however, the device needs to be operated at room temperature with a manageable, preferably non-magnetic, actuation and detection system. Here, we demonstrate a silicon nanomechanical beam structure operating as a two-state system at room temperature with on-chip actuation and detection using standard electrostatic techniques.[9] In contrast to previous experiments[8] where switching is accomplished by applying a force (additional to the driving) here we induced switching by modulating solely the phase of the driving force.

In a typical static operation, a suspended structure forming one component of the variable capacitance moves from one fixed position (ON) to a second fixed position (OFF). In the dynamic configuration, the suspended part of the structure continuously vibrates at its resonance frequency or at a frequency within the nonlinear hysteretic regime. The respective ON and OFF configurations are defined by two distinct states, corresponding to two specific amplitudes of vibration. Switching between these two states is achieved using a modulation (trigger or write) signal. In contrast to varying the amplitude of the modulation, which results in a continuous change in amplitude, here we demonstrate switching as a function of the phase of the drive signal.

Our device is fabricated, using standard e-beam lithography and surface nanomachining[8], from single-crystal silicon using a silicon on insulator (SOI) wafer. Figure 1 includes a micrograph of the device (top view), including actuation and detection electrodes and the central beam, which is 15 μm long, 500 nm thick and 300 nm wide. The gap ($g$) between the beam and the actuation/detection electrodes is 300 nm. For actuation, the beam is biased with a dc voltage ($V_B$ = 12 V for all results in this letter) and a megahertz-frequency voltage is applied to one of the side electrodes using a network analyzer or a signal generator, as shown in Fig. 1. This produces an in-plane force of magnitude

$$F(t) = \frac{1}{2}\frac{dC_1}{dx}(V_B + V_D\cos(\omega t))^2 + \frac{1}{2}\frac{dC_2}{dx}V_B^{\ 2}, \qquad (1)$$

where $C_1$ ($C_2$) is the capacitance between the beam and the excitation (detection) electrode, $V_D$ is the drive amplitude, and $x$ is the effective displacement of the beam. Assuming $\frac{d^n C_1}{dx^n} = (-1)^n \frac{d^n C_2}{dx^n}$ for parallel plate capacitor configurations, and expanding this expression in terms of $\frac{V_D}{V_B}$ and $\frac{x}{g}$ (where $V_B >> V_D$ and $g >> x$) around the equilibrium position ($x$=0) one obtains

$$F(t) \approx C' V_B V_D \cos(\omega t) + V_B{}^2 (C'' x + \frac{1}{6} C^{(IV)} x^3), \qquad (2)$$

where $C' = \left.\frac{dC_1}{dx}\right|_{x=0} = -\left.\frac{dC_2}{dx}\right|_{x=0}$ and so on.. The oscillating beam produces a time varying capacitance between the beam and the detection electrode, which in the presence of a constant potential creates a current $i = \frac{dQ}{dt} = V_B \frac{dC_2}{dt} = V_B \frac{dC_2}{dx} \dot{x} \approx -V_B C' \dot{x}$ . This current is amplified using a transimpedance amplifier and measured using a network analyzer.

For small drive amplitudes, the expected response as a function of drive frequency is the standard Lorentzian line shape, characteristic of the linear regime with a resonance frequency of $\frac{\omega_0}{2\pi} = 4.7$ MHz and a quality factor, Q $\approx 100$ (at $10^{-3}$ torr). It is important to note that the value of the loaded Q is limited by the electrical dissipation; hence decreasing the pressure below 1 millitorr does not improve the quality factor (as is expected to for an equivalent, unloaded resonator). As the drive amplitude is increased, the beam enters the nonlinear regime. Fig. 2 (a) shows the nonlinear response of the beam as a function of the drive amplitude for a fixed frequency of 4.83 MHz. The response exhibits a range of amplitudes in which the beam is bistable. In the nonlinear regime, the beam can be described by the Duffing equation with on degree of freedom[10],

$$\ddot{x} + 2\gamma \dot{x} + \omega_0^2 x + k_3 x^3 = f(t), \qquad (3)$$

where $\gamma$ (dissipation coefficient), $\omega_0^2$ and $k_3$ (nonlinear coefficient) are functions of the bias voltage, and *f(t)* represents the force component with explicit time dependence (see Eq. 2). In our case, the frequency-response curve bends towards higher frequencies with increasing drive amplitude consistent with $k_3 > 0$ .

Using the setup described in Fig. 1, a voltage $V = V_B + V_D \cos(\omega t + \frac{\varphi_0}{2}\Theta(\Omega))$ is applied on the excitation electrode, where $\varphi_0$ is the phase deviation and $\Theta(\Omega)$ represent a square wave of period $\frac{2\pi}{\Omega}$ ($\Theta(\Omega)$ = 1 for the first half of the period and -1 for the other half). This voltage produces a force per unit mass $f(t) = f_D \cos(\omega t + \frac{\varphi_0}{2}\Theta(\Omega))$ where the values of $\omega$ is chosen to lie in the bistable region. Using setup (I) (see Fig 1), the bistable region is measured by sweeping the drive amplitude with a very large modulation period. Subsequently, the amplitude is fixed to a given point (broken line in Fig 2 (a)) and the period of modulation increased (to 2 seconds in this example). The result is presented in Fig. 2 (b) where one observes discrete jumps in the beam amplitude, corresponding to two distinct states (upper graph), as a consequence of the phase modulation (lower graph). These jumps are synchronized with the modulating signal and the values of the two amplitude states correspond to the measurements in the hysteresis curve. This phase modulation induced switching is in contrast to the static additive force previously used.[8] Only including a single signal resulted in a more stable electric circuit response, hence increasing the control over the switching events and the ease to induce them.

Setup (II), illustrated in Fig 1, is used to study the dependence of the switching as a function of the phase deviation where a signal generator is used to drive the beam. This allows us to controllably vary the phase deviation. We observe switching events for values of the phase deviation ranging from 0.9 π to 1.75 π . To measure the fidelity the switching fraction is shown in Fig. 3 (a). This quantity is defined as the number of amplitude switches divided by the number of phase switches. In Fig. 3 (b) three time series of switching events are shown for different phase deviations (inset Fig. 3(a)). One can see that when the phase deviation in not large enough the beam skips periods, but the successful switches are synchronized with the modulation. For all points in Fig. 3, the modulation frequency $\frac{\Omega}{2\pi}$=10 Hz. No spontaneous switching was observed.

An important feature of a memory (switching) element is the read-write frequency. While the read frequency is given by the resonance frequency with a bandwidth defined by the loaded quality factor, the writing speed is given by the switching frequency. Due to the experimental setup utilized here, switching only up to ~1 kHz has been properly measured. For higher speeds, the noise level becomes comparable to the jump size. In theory, the open-loop speed is given by f/Q ~ 48 kHz.

It is noteworthy that all the observed features of the switching mechanism can be obtained at least qualitatively by simple numerical integration of Equation 3. It is well known that the hysteresis curve (Fig. 2 (a)) can be easily reproduced.[10] However, by modulating the phase of the drive one can also obtain switching events with a phase deviation dependency similar to the one reported here (Fig. 3).

In conclusion, we demonstrate a fully controllable room-temperature nanomechanical switching element, actuated and sensed using standard electrostatic techniques. Implementing a novel phase modulation scheme, we show that the two states in the hysteretic nonlinear regime of a nanomechanical resonator can be controlled with 100% fidelity. This silicon-based switching device can be fabricated with on-chip CMOS circuitry to provide unprecedented advantages of size and integration. This work is supported by the National Science Foundation (Career Grant No. DMR 0449670).

**References:**

*a) E-mail: mohanty@buphy.bu.edu*

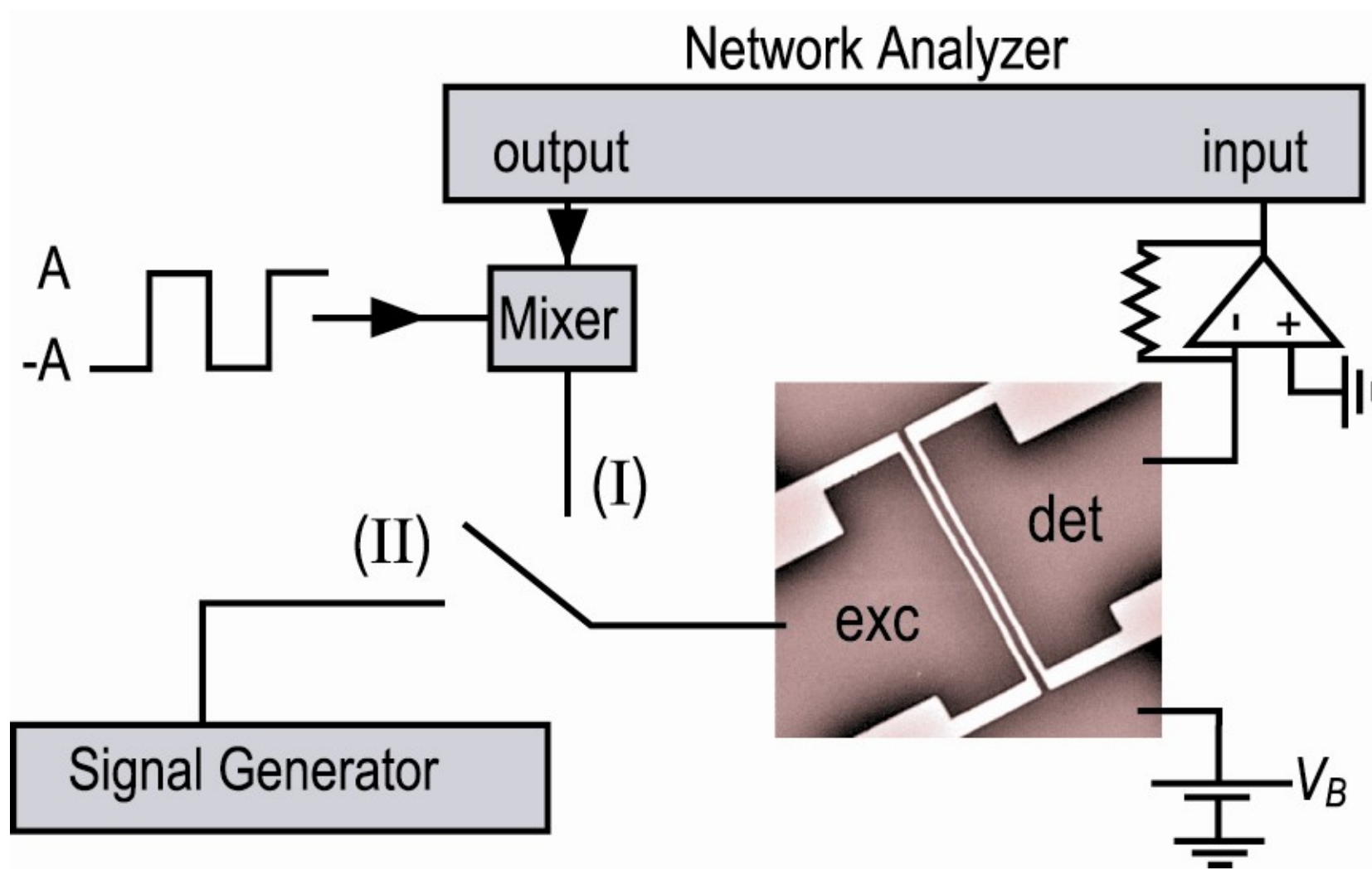

**FIG. 1.** (Color online) Experimental setup. In configuration (I), the resonator is excited by the output of a vector network analyzer (Agilent N3383A), which, mixed with a square wave, produces a phase modulation with phase deviation π. In configuration (II), the signal generator (Agilent 33220A) is used to excite the beam with a variable phase deviation. The output current is amplified by a transimpedance amplifier and measured with the vector network analyzer set to continuous wave (CW) time mode (measures the time dependence of the beam amplitude and phase at the excitation frequency). The electrodes are named excitation (exc) and detection (det) for reference.

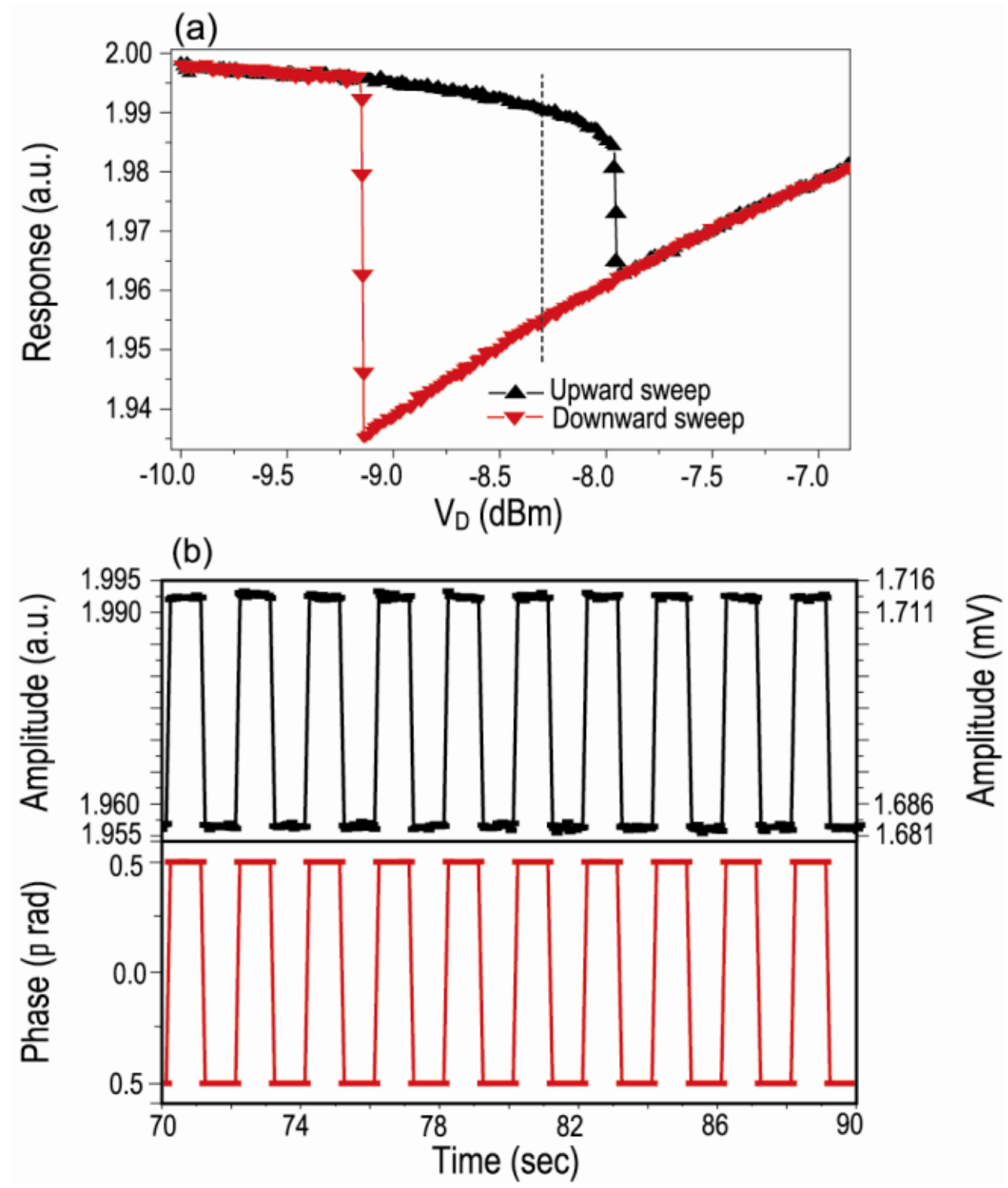

**FIG 2.** (Color online) (a) Amplitude hysteresis for a constant drive frequency of 4.83 MHz. (b) (top) Switching events. CW time measurement of the beam at the pump frequency shows the switching between two states in response to a square wave modulation of the drive phase (bottom). The drive amplitude is -8.3 dBm (broken line in (a)). The switching events are synchronized with the modulation, and their size corresponds to the jump height on the hysteresis curve. The data set is taken with setup (I)**.**

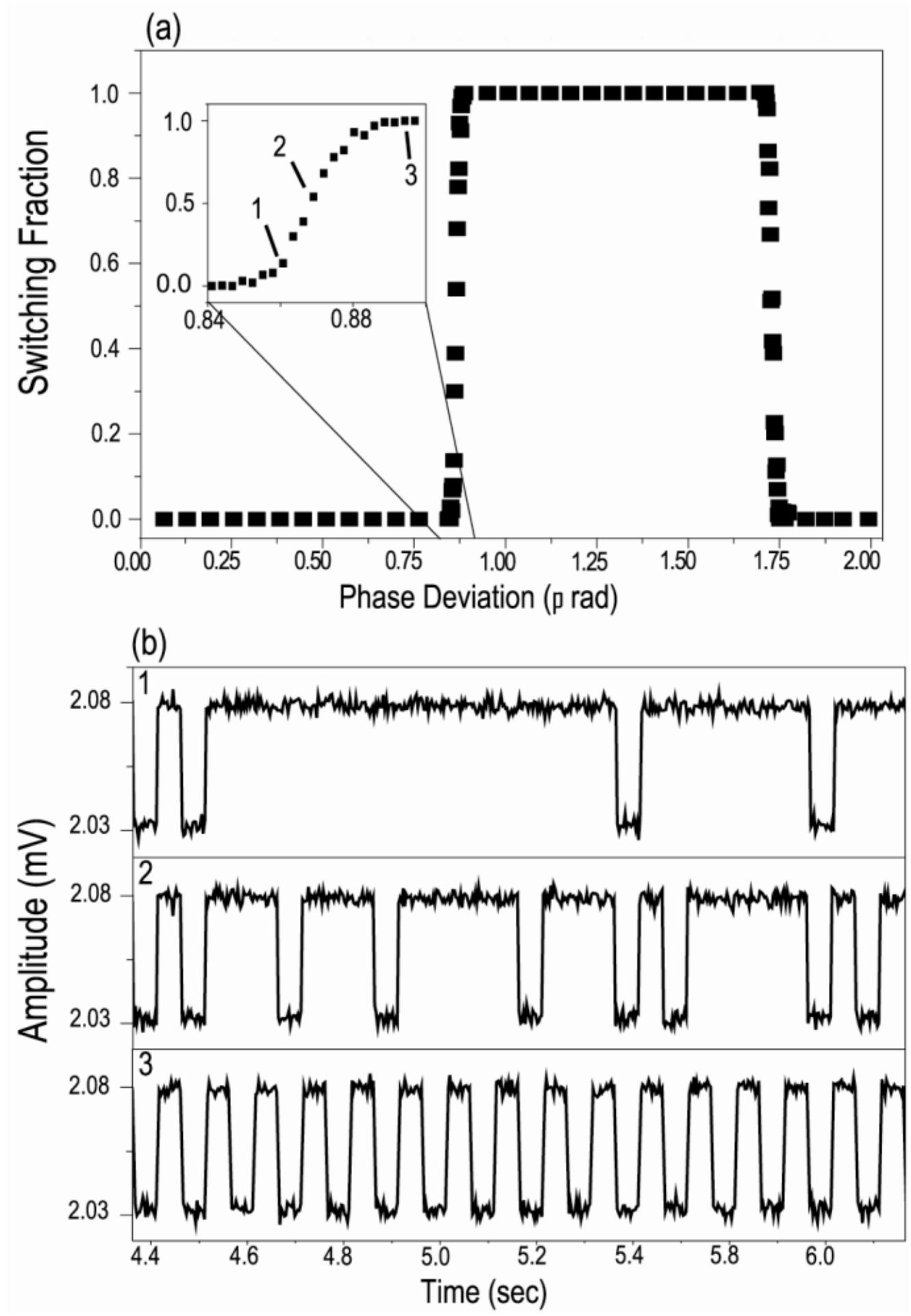

**FIG 3.** (a) Switching fraction as a function of the phase deviation. For all these points the driving power is -9 dBm at 4.9 MHz. (b) Response of the beam for different phase deviations of the modulation (marked in the inset of (a)). Graph 1 has a phase deviation of 0.860, graph 2 0.867 and graph 3 0.887 $\pi$ radians. The modulation frequency is 10 Hz for the three graphs. The data set is taken using setup (II).